# Spin interference in silicon one-dimensional rings

**N T Bagraev[1], N G Galkin[1], W Gehlhoff[2], L E Klyachkin[1], A M Malyarenko[1] and I A Shelykh[3]**
[1]Ioffe Physico-Technical Institute, RAS, 194021 St. Petersburg, Russia
[2]Institut für Festkörperphysik, TU Berlin, D-10623 Berlin, Germany
[3]Physics and Astronomy School, University of Southampton, Highfield, Southampton, UK

E-mail:

**Abstract.** We present the first findings of the spin transistor effect in the Rashba gate-controlled ring embedded in the p-type self-assembled silicon quantum well that is prepared on the n-type Si (100) surface. The coherence and phase sensitivity of the spin-dependent transport of holes are studied by varying the value of the external magnetic field and the bias voltage that are applied perpendicularly to the plane of the double-slit ring. Firstly, the amplitude and phase sensitivity of the *$0.7 \cdot (2e^2/h)$* feature of the hole quantum conductance staircase revealed by the quantum point contact inserted in the one of the arms of the double-slit ring are found to result from the interplay of the spontaneous spin polarization and the Rashba spin-orbit interaction. Secondly, the quantum scatterers connected to two one-dimensional leads and the quantum point contact inserted are shown to define the amplitude and the phase of the Aharonov-Bohm and the Aharonov-Casher conductance oscillations.

The spin-correlated transport in low-dimensional systems was in focus of both theoretical and experimental activity in the last decade [1,2]. The studies of the Rashba spin-orbit interaction (SOI) that results from the structure inversion asymmetry in mesoscopic nanostructures have specifically attracted much of the efforts [1,3,4]. The variations in the Rashba SOI value appeared to give rise to the developments of spintronic devices based on the spin-interference phenomena that are able to demonstrate the characteristics of the spin field-effect transistor (FET) even without ferromagnetic electrodes and external magnetic field [4,5]. For instance, the spin-interference device shown schematically in figure 1a represents the Aharonov-Bohm (AB) ring covered by the top gate electrode, which in addition to the geometrical Berry phase provides the phase shift between the transmission amplitudes for the particles moving in the clockwise and anticlockwise direction [4]. This transmission phase shift (TPS) seems to be revealed by the Aharonov-Casher oscillations of the conductance measured under the top gate voltage applied to the two-terminal device, with the only drain and source constrictions [5]. However, the variations in the density of carriers that accompany the application of the top gate voltage are also able to result in the conductance oscillations. Therefore the three-terminal device with the quantum dot (QD) [6] or the quantum point contact (QPC) [7] inserted in one of the ring's arms using the split-gate technique [8] could be more appropriate to divide the relative contribution of the Aharonov-Casher (AC) effect in the conductance oscillations (see figure 1a).

Since the AB ring's conductance has to be oscillated with the periodicity of a flux quantum h/e when a variable magnetic field threads its inner core, these AB oscillations have been shown to be persisted, if the transport through the QD [6] or QPC [7] inserted is coherent. The TPS caused by the QD or QPC has been found to be equal to π in the absence of the spin polarisation of carriers [6,7], whereas the TPS value π/2 appeared to be very surprisingly in the range of the *0.7·(2e²/h)* feature of the quantum conductance staircase [9] revealed by the QPC inserted thereby verifying the spin polarisation in the AB ring [7]. This TPS is of interest to cause no any changes in the amplitude of the *0.7·(2e²/h)* feature, thus remaining unsolved the question on the relative contribution of the spontaneous spin polarization and the Rashba SOI to its creation. Nevertheless, the high sensitivity of the *0.7·(2e²/h)* feature to the variations of the spin polarisation of carriers allowed the three-terminal device with the QPC inserted to study the AC conductance oscillations.

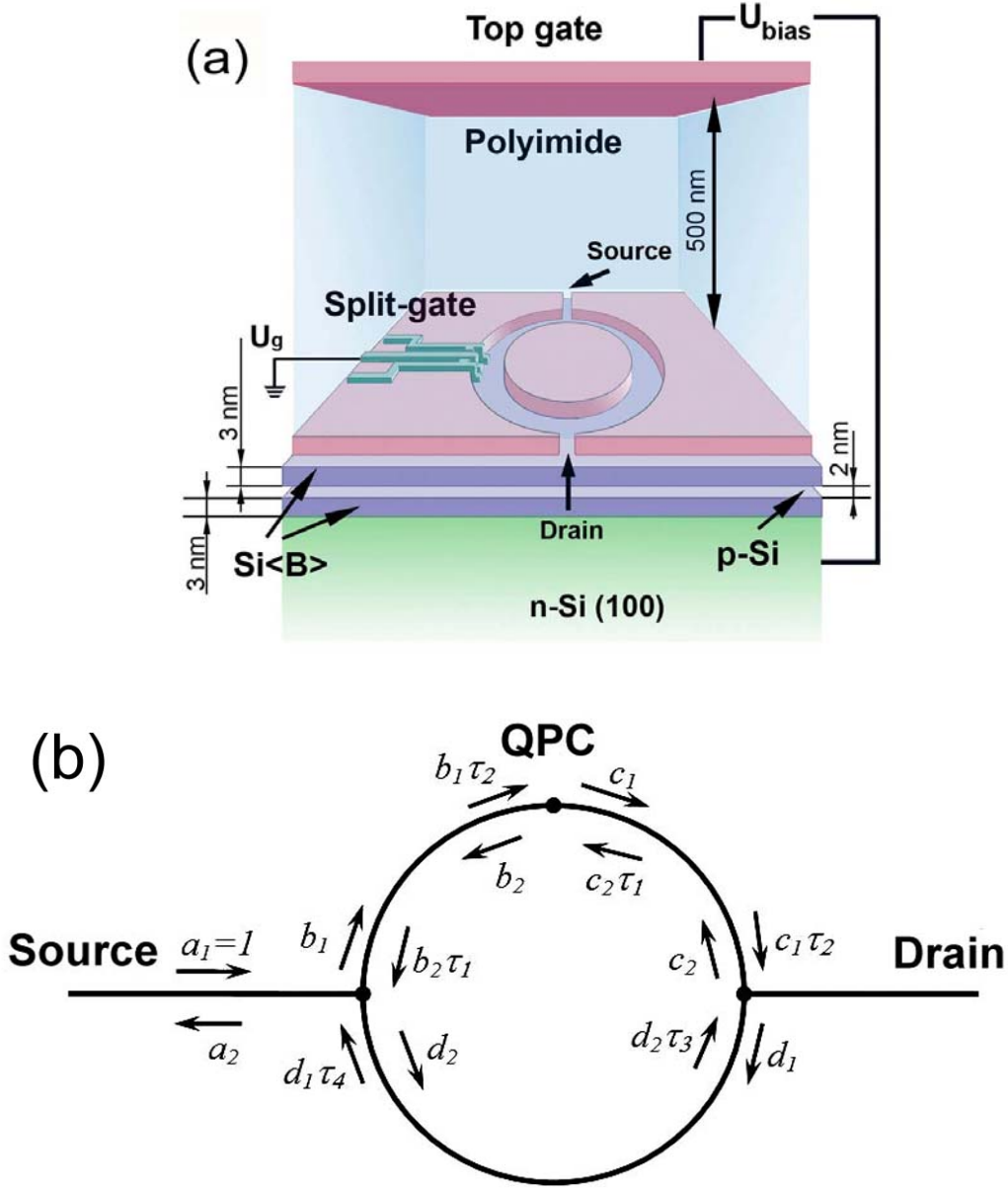

**Figure 1.** (a) Schematic diagram of the device that demonstrates a perspective view of the p-type silicon quantum well located between the δ - barriers heavily doped with boron on the n-type Si (100) surface, the top gate that is able to control the sheet density of holes and the depletion regions created by split-gate method, which indicate the double-slit ring with QPC inserted in one of its arms. (b) Schematic view of a spin-interference device based on the AB ring with the QPC inserted in its arm, which is connected with two one-dimensional leads by the QPCs and covered by the gate electrode that controls the Rashba SOI with the amplitudes of travelling electronic waves.

Here we report the measurements of the amplitude and the TPS of the $0.7 \cdot (2e^2/h)$ feature by tuning the Rashba SOI in the three-terminal silicon one-dimensional ring (figure 1a). The device advanced firstly in Refs [7, 10] is based on the ultra-narrow, 2 nm, self-assembled silicon quantum well (Si-QW) of the p-type that is prepared on the n-type Si (100) surface by the short-time diffusion of boron from the gas phase. The one-dimensional ring embedded electrostatically in the Si-QW, R=2500 nm, contains the source and the drain constrictions that represent QPCs as well as the QPC inserted in its arm by the split-gate method.

The parameters of the high mobility Si-QW were defined by the Hall measurements as well as by the SIMS, STM cyclotron resonance and EPR methods. The initial value of the sheet density of 2D holes, $4 \cdot 10^{13}$ m$^{-2}$, was changed controllably over one order of magnitude, between $5 \cdot 10^{12}$ m$^{-2}$ and $9 \cdot 10^{13}$ m$^{-2}$, by biasing the top gate above a layer of insulator, which fulfils the application of the p$^+$-n bias junction. The variations in the mobility measured at 3.8 K that corresponded to this range of the $p_{2D}$ values appeared to occur between 80 and 420 m$^2$/vs. Thus the value of the mobility was high even at low density of the 2D holes. Besides, the high value of mobility showed a decrease no more than two times in the range of temperatures from 3.8 K to 77 K that seems to be caused by the ferroelectric properties for the δ – barriers heavily doped with boron [11], which confine the Si-QW (figure 1a).

These characteristics of the 2D gas of holes allowed the studies of the quantum conductance staircase revealed by the heavy holes at 77 K (figure 2). The number of the highest occupied mode of the QPC inserted in the arm of the AB double-slit ring was controlled by varying the split-gate voltage, whereas the Rashba SOI was tuned by biasing the top gate. The experiments are provided by the effective length of the QPC inserted, 0.2 μm, and the cross section of the 1D channel, 2 nm × 2 nm, which is determined by the width of the Si-QW and the lateral confinement due to ferroelectric properties for the δ – barriers. We focus then on the effects of the bias voltage controlled by the top gate and the external magnetic field on the amplitude and the TPS of the $0.7 \cdot (2e^2/h)$ feature at the fixed value of the split-gate voltage, 5.7 mV, in its range (see figure 2).

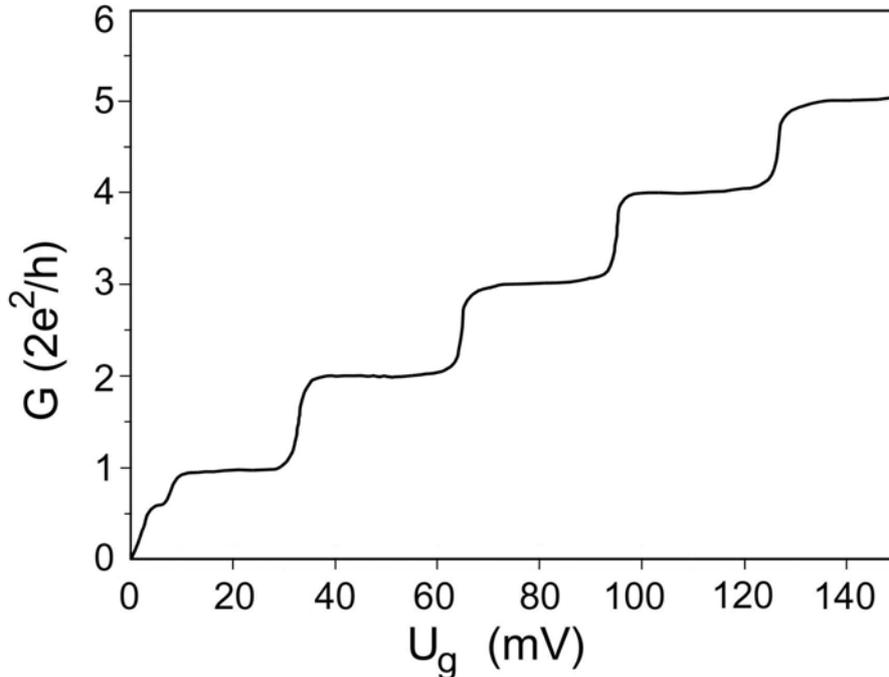

**Figure 2.** The quantum conductance staircase as a function of the split-gate voltage, which is revealed by heavy holes tunneling through the QPC inserted in the arm of the Si-based double-slit ring in zero magnetic field and under zero bias voltage controlled by the top gate (see figure 1a).

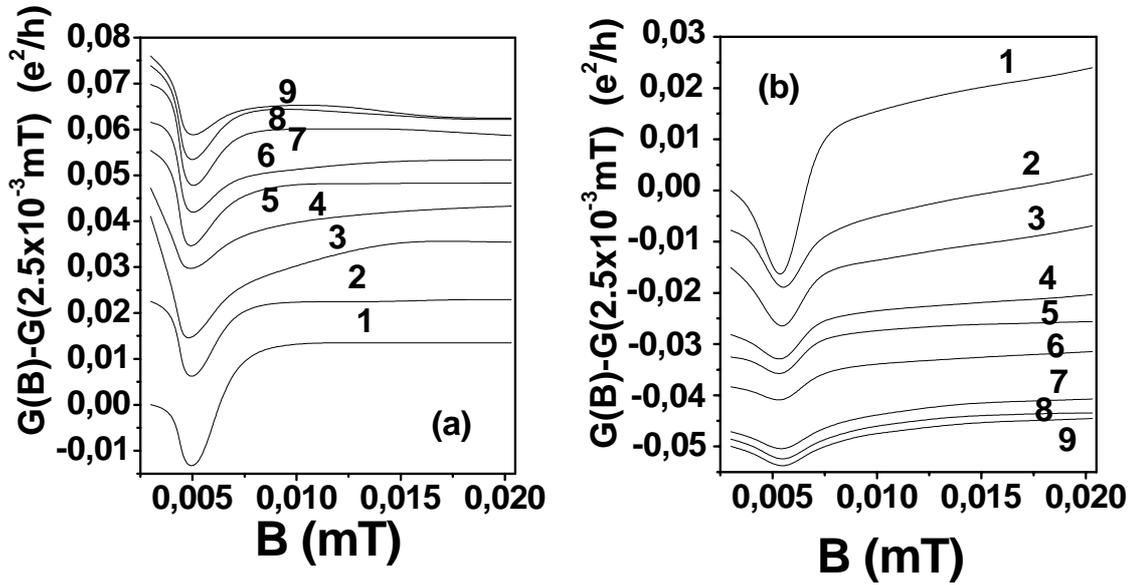

**Figure 3.** Experimental magnetoconductance, $G(B)-G(2.5\cdot10^{-3}mT)$, for different negative (a) and positive (b) $U_{bias}$ determined by the top gate voltage. $B=2.5\cdot10^{-3}$mT is the residual magnetic field obtained after screening the Earth magnetic field. A crossover from the weak localization to the weak antilocalization is revealed by varying the top gate voltage from negative to positive $U_{bias}$.
(a) $U_{bias}$, mV: 1 – 40, 2 – 140, 3 – 230, 4 – 260, 5 – 300, 6 – 330, 7 – 370, 8 – 390, 9 – 400.
(b) $U_{bias}$, mV: 1 – 30, 2 – 90, 3 – 140, 4 – 230, 5 – 260, 6 – 300, 7 – 360, 8 – 370, 9 – 380.

The presence of the Rashba SOI in the device studied is revealed by the measurements of the magnetoconductance (figures 3a and 3b). By varying the top gate voltage at the fixed value of the split-gate voltage, 5.7 mV, the transition from the positive magnetoresistance to the negative magnetoresistance is observed thereby verifying a crossover from the weak antilocalization to the weak localization following the changes of the concentration of the 2D holes. The dependencies in figures 3a and 3b are in a good agreement with the theoretical predictions [12] and correlate with the experimental electron magnetoconductance data in a high mobility $In_xGa_{1-x}As/InP$ quantum well [13]. The crossover from the weak antilocalization to the weak localization demonstrates a possibility of the SOI control by changing the value and sign of $U_{bias}$. It should be noted that in the range of extremely sheet density of 2D holes the opposite transformation from the negative to the positive magnetoresistance is found demonstrating the saturation of the magnetoconductance with the $p_{2D}$ concentration (see figure 3b, curves 7-9). This effect seems to be caused the enhancement of the spin-lattice relaxation time in ultra-narrow quantum wells. Thus, the SOI effects observed in the studies of the magnetoconductance are evidence of the constructive and destructive backscattering associated with QPCs by varying the top gate voltage thereby allowing the findings of the AC conductance oscillations in the range of the external magnetic fields outside the region of the weak antilocalization.

Figures 4a and 4b show the changes in the amplitude of the *0.7·(2e²/h)* feature found by biasing the top gate that appeared to be followed by increasing and decreasing the concentration of 2D holes from the initial value, $4 \cdot 10^{13}$ m$^{-2}$, with applying respectively the reversal and forward bias. At low densities of the 2D holes (see figure 4b), the amplitude of the *0.7·(2e²/h)* feature reduces under the forward bias up to the *0.7·(e²/h)* value crossing the point of *0.5·(2e²/h)* that indicates the spin degeneracy lifting for the first step of the quantum conductance staircase [9]. These data seem to be evidence of the spontaneous spin polarization of heavy holes in the 1D channel that is due to the efficient quenching of the kinetic energy by the exchange energy of carriers [10,14]. However, the amplitude of the *0.7·(2e²/h)* feature lower than *0.5·(2e²/h)* value, which has been also observed in Refs [15,16], is puzzling and needs to be studied in detail. In one's turn, figure 4a demonstrates the evolution of the *0.7·(2e²/h)* feature with increasing $p_{2D}$ from the *0.6·(2e²/h)* value at zero bias voltage applied to the top gate to *0.75·(2e²/h)* value under the reversal bias. These changes seem to be caused by increasing the linear concentration of heavy holes that attains the critical value corresponding to the spin depolarization in the 1D channel [10].

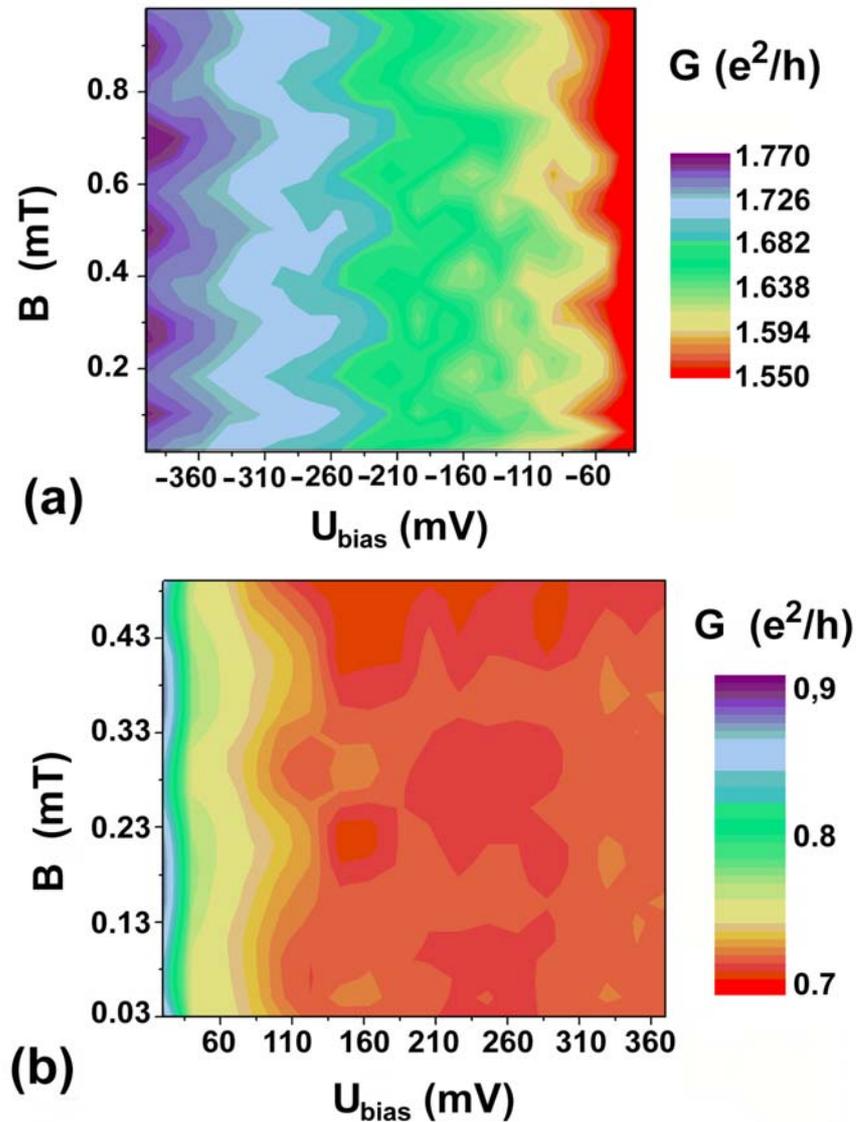

**Figure 4.** Plots of *G* vs *B* and $U_{bias}$ applied to the experimental device that reveal both the Aharonov-Bohm and the Aharonov-Casher conductance oscillations caused by the variations of the external magnetic field and the bias voltage controlled by the top gate, respectively. (a) reversal bias voltage; (b) forward bias voltage.

The amplitude of the *0.7·(2e²/h)* feature fixed by the split-gate voltage is also found to exhibit the Aharonov-Casher (AC) conductance oscillations that are due to the Rashba SOI by varying the bias voltage applied to the top gate (figures 4a and 4b). Figure 4a demonstrates the interplay of the AC conductance oscillations with the AB oscillations in all range of the reversal bias voltage and relatively weak magnetic fields, which indicates the principal role of the Rashba SOI in their formation. The gradual smoothing of the conductance oscillations of both types with decreasing the density of the 2D holes under the forward bias appears to be due to the localization of the heavy holes that is accompanied by the spin-lattice relaxation processes (figure 4b) [7,10].

These phase variations of the AB and AC oscillations observed seem to be caused by the elastic scattering of the heavy holes on the QPCs inside the double-slit ring. Therefore we used the scattering matrix approach to model theoretically the experimental results [17]. In frameworks of ballistic regime, the spin-dependent conductance of the three-terminal device (figure 1b) was calculated as

$$G_{ds\pm} = \frac{e^2}{h}|f_{ds\pm}|^2 \qquad (1)$$

where the index $\pm$ corresponds to the spin projection of the carrier on its wavenumber in the outgoing leads, $f_{ds\pm}$ are corresponding transmission amplitudes. As above noticed, the magnitude of conductance, $G_{ds\pm} = G_{ds+} + G_{ds-}$, and the spin polarization of the outgoing current are determined by the phase relations between the waves propagating in the AB ring in the clockwise and anticlockwise directions, which are controlled by varying the Rashba parameter $\alpha$ as well as the value of the external magnetic field and the Fermi energy, $E_F$. The Rashba parameter for the heavy holes depends linearly on the value of the perpendicular electric field $E_z$ and reads

$$\alpha_{hh} = -3\beta_{hh}\langle k_r^2 \rangle E_z \qquad (2)$$

where $\langle k_r^2 \rangle$ is an average value of the square of the particle's wavenumber in the plane perpendicular to the axis of the wire, the constant $\beta_{hh}$ depends on the details of the band structure of the material [18,19].

In frameworks of the model presented, the case of the zero temperature is accented to provide the step-liked energy distribution of the carriers. Besides, the potentials of the two outgoing leads $V_{ds}$ are taken to be equal and small enough, $eV_{ds} \ll E_F$, so that the only carriers whose energy lies in the vicinity of the Fermi surface participate in the transport. The radius of the AB ring is taken to be much smaller than inelastic scattering length to satisfy the conditions of the ballistic transport. Besides, the conjunctions between the AB ring and leads are modelled by the quantum point contacts (QPCs) which provide the elastic scattering of the carriers. The QPCs are presumed to be identical and spin-independent. The latter assumption means that the spin of the carrier conserves during the passing through the QPCs. Each QPC is characterised by the amplitude of the elastic backscattering of the carrier propagating inside the lead, $\sigma$, $|\sigma|<1$, that is determined by the system geometry [20]. The QPCs become completely transparent, if $\sigma=0$. These conditions allow us to use the Landauer-Buttiker approach for the calculations of the conductance $G_{ds\pm}$ [21,22].

The conductance of the three-terminal ring results from the phaseshifts of the waves propagating between the different QPCs (see figure 1b)

$$\tau_1 = \exp\left[i\frac{\pi}{2}\left(k_+a - \Phi/\Phi_0 + \frac{1}{2}\right)\right] \qquad (3a)$$

$$\tau_2 = \exp\left[i\frac{\pi}{2}\left(k_-a + \Phi/\Phi_0 - \frac{1}{2}\right)\right] \qquad (3b)$$

$$\tau_3 = \exp\left[i\pi\left(k_+ a - \Phi/\Phi_0 + \frac{1}{2}\right)\right] \tag{3c}$$

$$\tau_4 = \exp\left[i\pi\left(k_- a + \Phi/\Phi_0 - \frac{1}{2}\right)\right] \tag{3d}$$

Where $\Phi = \pi a^2 B$ is the magnetic flux through the AB ring, as well as $a$ is the radius of the AB ring and $\Phi_0$ is an elementary flux quantum. The first term at the relationships corresponds to the Aharonov-Casher phase, the second to the Aharonov-Bohm phase, and the third to the geometrical Berry phase [4]. Finally, mutual orientation of the spin and the effective magnetic field is opposite to the discussed above, when the spin of the carrier in the ingoing lead is opposite to its wave vector. In this case, the values of $k_+$ and $k_-$ should be interchanged in the calculations of the phaseshifts [23]:

$$k_\pm = \pm\frac{m\alpha}{\hbar^2} + \sqrt{\frac{m}{\hbar^2}\left(\frac{m\alpha}{\hbar^2} + 2E_F\right)} \tag{4}$$

The phase shifts that are seen to be dependent on the Rashba parameter $\alpha$ determined by the effective magnetic field, which is created by the Rashba SOI, $\mathbf{B}_{eff} = \frac{\alpha}{g_B\mu_B}[\mathbf{k}\times\mathbf{e}_z]$, and consequently on the modulation of the conductance, are observed not only in the AB oscillations as a function of the magnetic field, but also in the Aharonov-Casher (AC) oscillations [4,5,17] as a function of the bias voltage controlled by the top gate. The equations (2)-(4) lead to the analytical expressions for the amplitudes of the transmission into the two outgoing leads. The details of the calculation can be found in the Ref 24.

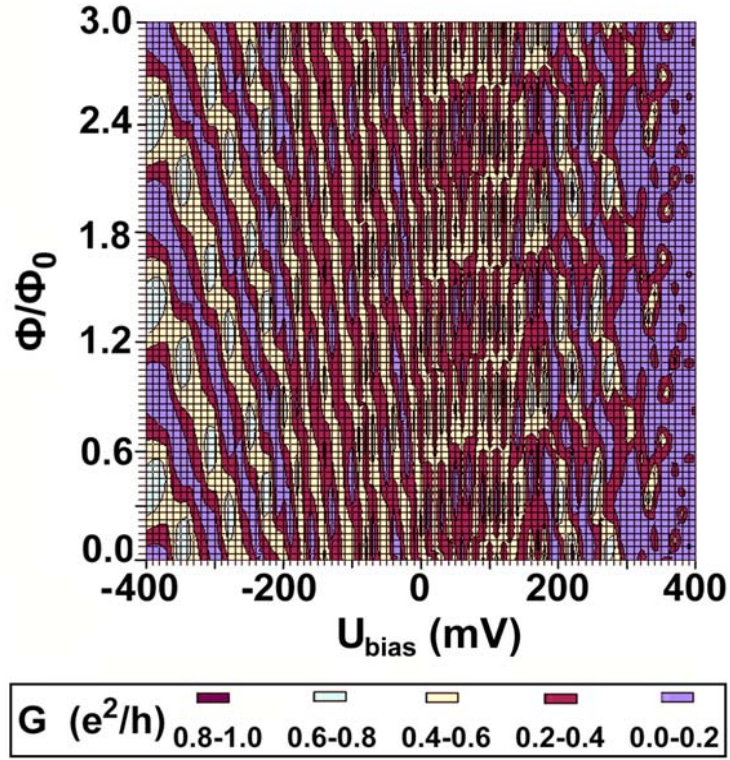

**Figure 5.** Plots of $G$ vs $B$ and $U_{bias}$ calculated using the parameters of the experimental device, which reveal both the Aharonov-Bohm and the Aharonov-Casher conductance oscillations caused by the variations of the external magnetic field and the bias voltage controlled by the top gate, respectively.

The results of the numerical calculations of the conductance as a function of the bias voltage and the external magnetic field are present in figure 5. Both AB and AC oscillations are clearly seen, thereby verifying the principal role of the Rashba SOI in the spin-dependent transport in the three-terminal device. Specifically, decreasing the density of the 2D holes under the forward bias shows the gradual quenching of the conductance oscillations that is due to the localization of carriers.

In conclusions, we have demonstrated experimentally that tuning the Rashba coupling parameter induced by the bias voltage applied to the AB ring with three asymmetrically situated electrodes, it is possible to control efficiently its conductance in the ballistic regime by the measurements of the AC oscillations. The interplay between the Rashba SOI and spontaneous spin polarization of carriers that gives rise to their relative contributions to the spin-dependent transport phenomena in the three-terminal one-dimensional rings have been also established.

**References**
[1] Rashba E I 2000 *Phys. Rev.* **B 62** R16267
[2] Wolf S A, Awshalom D D, Buhrman R A, Daughton J M, von Molnar S, Roukes M L, Chtchelkanova A Y and Treger D M 2001 *Science* **294** 1488
[3] Bychkov Yu A and Rashba E I 1984 *J.Phys.* **C 17** 6039
[4] Aronov A G and Lyanda-Geller Y B 1993 *Phys. Rev. Lett.* **70** 343
[5] Nitta J, Meijer F E and Takayanagi H 1999 *Appl. Phys. Lett.* **75** 695
[6] Schuster R, Buks E, Heiblum M, Mahalu D, Umansky V and Shtrikman H 1997 *Nature* **385** 417
[7] Bagraev N T, Bouravleuv A D, Gehlhoff W, Ivanov V K, Klyachkin L E, Malyarenko A M, Rykov SA and Shelykh I A 2002 *Physica* **E 12** 762
[8] Thornton T J, Pepper M, Ahmed H, Andrews D and Davies G J 1986 *Phys. Rev. Lett.* **56** 1198
[9] Thomas K J, Nicholls J T, Simmons M Y, Pepper M, Mace D R and Ritchie D A 1996 *Phys. Rev. Lett.* **77** 135
[10] Bagraev N T, Shelykh I A, Ivanov V K and Klyachkin L E 2004 *Phys. Rev.* **B 70** 155315
[11] Bagraev N T, Bouravleuv A D, Klyachkin L E, Malyarenko A M, Gehlhoff W, Ivanov V K and Shelykh I A 2002 *Semiconductors* **36** 439
[12] Miller J B, Zumbuehl D M, Marcus C M, Lyanda-Geller Y B, Goldhaber-Gordon D, Campman K and Gossard A C 2003 *Phys. Rev. Lett.* **90** 076807
[13] Studenikin S A, Goleridze P T, Ahmed N, Poole P J and Sachrajda 2003 *Phys. Rev.* **B 68** 035317
[14] Ghosh A, Ford C J B, Pepper M, Beere H E and Ritchie D A 2004 *Phys. Rev. Lett.* **92** 116601
[15] Liang C–T, Pepper M, Simmons M Y, Smith C G and Ritchie D A 2002 *Phys. Rev.* **B 61** 9952
[16] Pyshkin K S, Ford C J B, Harrell R H, Pepper M, Linfield E H and Ritchie D A 2000 *Phys. Rev.* **B 62** 15842
[17] Shelykh I A, Bagraev N T, Galkin N G and Klyachkin L E 2005 *Phys. Rev.* **B 71** 113311
[18] Winkler R 2000 *Phys. Rev.* **B 62** 4245
[19] Winkler R 2002 *Phys. Rev.* **B 65** 155303
[20] Taniguchi T and Buttiker M 1999 *Phys. Rev.* **B 60** 13814
[21] Landauer R 1970 *Philos. Mag.* **21** 863
[22] Buttiker M. 1986 *Phys. Rev. Lett.* **57** 1761
[23] Altshuler B L, Aronov A G and Spivak B Z 1981 *Sov. Phys. JETP Lett.* **33** 94
[24] Shelykh I A, Galkin N G and Bagraev N T 2005 *Phys. Rev.* **B 72** 235316